\documentclass{osa-article}
\journal{oe}

\usepackage[utf8]{inputenc}

\usepackage{amsmath}
\usepackage{hyperref}
\usepackage{graphicx}
\usepackage{physics}
\usepackage{bbm}
\usepackage{url}
\usepackage{soul}
\hypersetup{
    colorlinks = true,
    linkcolor  = blue,
    citecolor  = blue,
    urlcolor   = blue,
}

\graphicspath{{Figures/}}

\newcommand{\trans}{\emph{\textsc{t}}} 
\newcommand{\nbar}{\bar{n}} 

\begin{document}

\title{Gaussian State-Based Quantum Illumination with Simple Photodetection}

\author{Hao Yang\authormark{1*}, Wojciech Roga\authormark{2}, Jonathan D. Pritchard\authormark{1}, and John Jeffers\authormark{1}}

\address{\authormark{1}Department of Physics, University of Strathclyde, Glasgow G4 0NG, U.K.\\
\authormark{2}National Institute of Information and Communications Technology (NICT), 4-2-1 Nukui-kita, Koganei, Tokyo 184-8795, Japan}

\email{\authormark{*}hao.yang@strath.ac.uk}

\begin{abstract}
	Proofs of the quantum advantage available in imaging or detecting objects under quantum illumination can rely on optimal measurements without specifying what they are. We use the continuous-variable Gaussian quantum information formalism to show that quantum illumination is better for object detection compared with coherent states of the same mean photon number, even for simple direct photodetection. The advantage persists if signal energy and object reflectivity are low and background thermal noise is high. The advantage is even greater if we match signal beam detection probabilities rather than mean photon number. We perform all calculations with thermal states, even for non-Gaussian conditioned states with negative Wigner functions. We simulate repeated detection using a Monte-Carlo process that clearly shows the advantages obtainable. 
\end{abstract}

\section{Introduction}
Quantum states of light for object detection in a noisy environment, coined as ``quantum illumination'', were originally introduced \cite{Lloyd2008a} and subsequently investigated for continuous variable Gaussian states \cite{Tan2008, UshaDevi2009, shapiro2009} - states with Gaussian Wigner functions. Quantum illumination uses quantum correlations to provide improved object detection when compared to classical light sources. The proof of this advantage, at optical frequencies or in quantum radar signal discrimination, boils down to an optimization problem focused on minimizing the probability of error in hypothesis testing; from thereon parameters are chosen to present a favourable picture where quantum states gain a performance advantage over illumination with coherent states. The advantage of entangled Gaussian states prevails in the lossy and noisy scenarios if both modes can be measured "optimally" \cite{Zhuang2017a} even if the entanglement is completely lost \cite{Zhang2013, zhang2015}. Then quantum correlations \cite{Roga2016a} may remain in the form of quantum discord \cite{Weedbrook2016}.

The problem can be stated in terms of quantum optical state discrimination, which is a well-studied subtopic in quantum information \cite{Pirandola2018, Barnett2009}. The goal is to detect successfully an object of weak reflectivity $\kappa$ possibly hidden in a strong thermal background noise. In effect we wish to determine which of two possible hypotheses, $H_0$ - the object is absent, or $H_1$ - the object is present, is true. If we send signal state $\hat{\rho}$ to a possible target then either $\hat{\rho}_0$ (corresponding to $H_0$ being true) or $\hat{\rho}_1$ corresponding to $H_1$) will return, conditioned by the absence or presence of the object. How distinguishable are these two states? Could we measure this difference? Previous analysis has circumvented what measurement we have to make and successive proofs are based on the Helstrom bound, which states that the absolute error in the optimal measurement is proportional to $1-\tfrac{1}{2}|\hat{\rho}_0-\hat{\rho}_1|$ \cite{Helstrom1969}. This is usually difficult to evaluate for Gaussian states, especially for states corresponding to multiple observation trials. Hitherto, a workaround solution has been to use easier-to-calculate fidelity bounds such as Uhlmann fidelity or the quantum Chernoff bound \cite{Fuchs1999, Jozsa1994, Marian2012, Audenaert2007} to estimate the error probability of the optimal measurement \cite{Sanz2017}. The quantum Chernoff bound provides the tightest approximation to the Helstrom bound under a large number of observations (see \cite{Weedbrook2012, Pirandola2008} for the Gaussian state analysis). However, this does not resolve the issue of identifying the exact measurement protocol required to perform discrimination.
\begin{figure}[t!]
    \centering
    \includegraphics[width=0.8\linewidth]{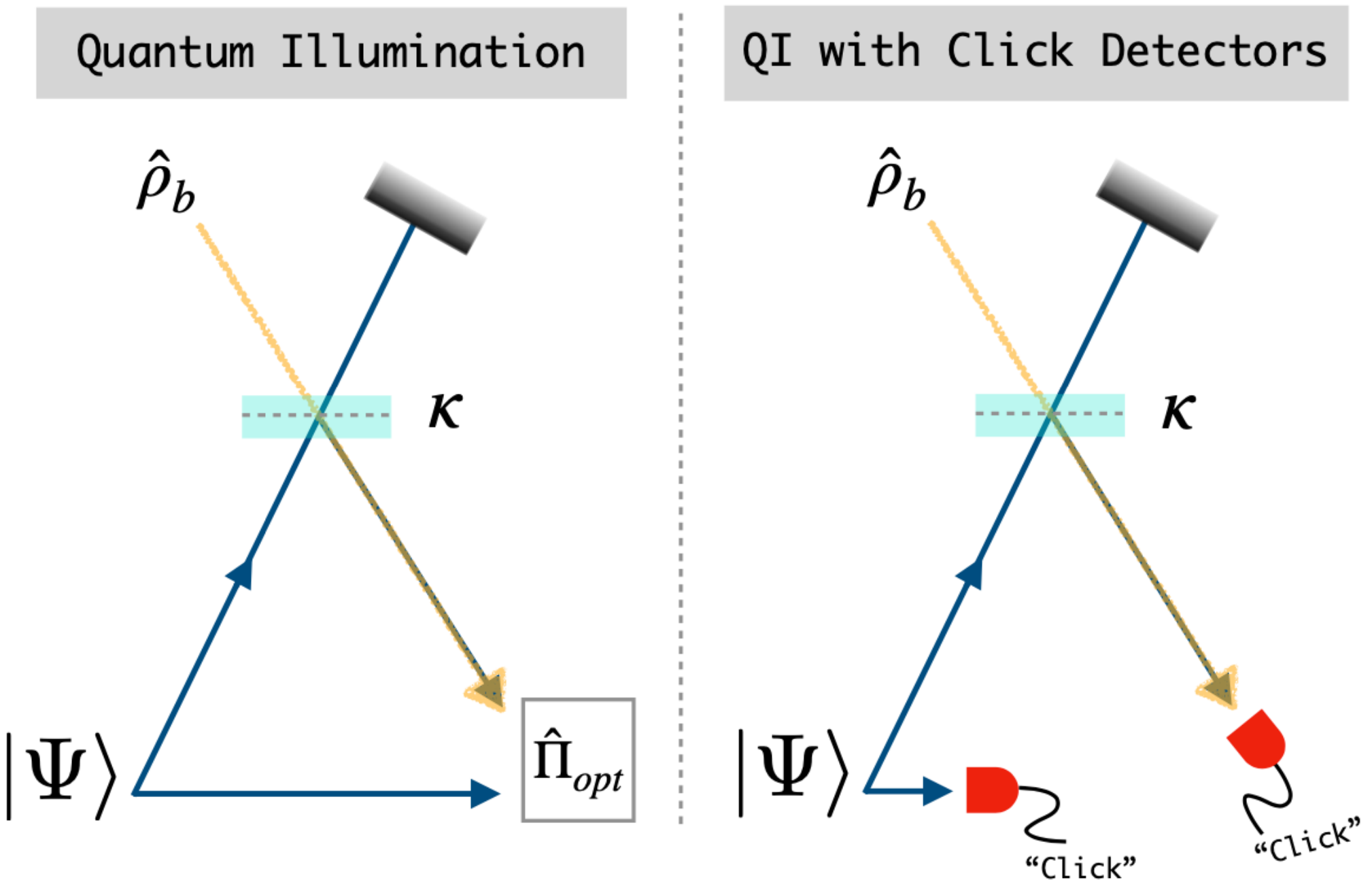}
    \caption{Optimal quantum illumination (QI) vs. QI with click detection. An object of reflectivity $\kappa$ is illuminated by one of a pair of quantum correlated beams $\ket{\Psi}$. The object mixes the reflected return signal with thermal background noise in state $\hat{\rho}_b$. The left panel shows the set up for optimal or joint detection, represented here by the operator $\hat{\Pi}_{opt}$. The right panel shows an explicit simpler detection method, where clicks from Geiger-mode photodetectors show enhanced click probabilities when the object is present.}
    \label{fig:QI_Schematic}
\end{figure}
Nevertheless, quantum illumination has been used and measured experimentally at optical frequencies with CCDs \cite{Lopaeva2013}, sum-frequency generating receivers \cite{Zhuang2017}, by fibre-coupled avalanche photodiodes \cite{England2019, Liu2019c}, and more recently in the microwave domain \cite{Guha2009,Barzanjeh2015, chang19, luong20, Barzanjeh2020}. Very recently, Ortolano et al \cite{Ortolano2020} have shown that photon-counting measurements with an entangled two-mode squeezed vacuum source can outperform any equal energy strategy based on statistical mixtures of coherent states in the quantum reading scenario \cite{Pirandola2011}.

In this paper we analyse an experimentally simpler approach to previous state discrimination methods for quantum illumination: a direct measurement strategy using positive operator-valued measures (POVMs) that model Geiger-mode photodetectors which can either fire or do not fire when EM radiation falls upon them or not respectively -- ``click'' or ``no-click'' (see Fig. \ref{fig:QI_Schematic} for difference in schematic set-up). Our approach uses entirely continuous-variable quantum information \cite{Braunstein2005, Ferraro2005, Weedbrook2012} with Gaussian states that are fully characterized by their statistical moments. In this way, some computational hurdles of discrete Fock-basis representations (large Hilbert spaces for multimode or higher energy states) are avoided  by working in the Gaussian-only domain. An essential point of the analysis is that it allows us to perform {\it all} calculations exactly within this Gaussian quantum optical framework, even though the states are not all Gaussian. Moreover, since this is a statistical analysis of QI, it is frequency independent and could therefore be applied both to lidar and to radar. By employing this simple click detection strategy, we can easily calculate the click probability of the return signal after its interaction with an external object, whilst incorporating any associated quantum efficiencies and thermal background noise sources, including detector dark counts. We compare results from two signal states with different photon-statistics: a coherent state that has classical photon-statistics and a quantum state heralded via click detection from the entangled two-mode squeezed vacuum (TMSV)\cite{Yang2020a}. As click-detection is not phase-resolving, the moments of the signal state, even after conditioning, can be parameterized by its average photon number $\nbar = \langle\hat{a}^\dagger\hat{a}\rangle$, aiding further simplification. Our purpose is to highlight where quantum illumination may provide an advantage over classical. Classical radar and lidar \cite{kuzmenko2020} are well developed technologies with many applications so quantum illumination will not be better in every situation, although some of its advantages for rangefinding have been pointed out recently \cite{frick2020}. We emphasize that there are states and measurement strategies that could give better results for object detection than we obtain, but these would typically require a local oscillator that is phase-locked to the signal. If the distance to the object is uncertain by more than a small fraction of a wavelength they are rendered impotent.

The paper is structured as follows. We give in Section 2 a brief introduction to continuous-variable bosonic Gaussian states and include some useful identities, before in Sections 3 and 4 introducing direct detection and the conditional states produced from performing a click-detection on one mode of the TMSV in a heralding process. In Section 5 we incorporate this into quantum illumination theory and compare results from coherent states and TMSV states. Specifically we calculate conditional probabilities of click detection for classical or quantum illumination in the presence or absence of a target object. We use these probabilities to calculate retrodictive conditional probabilities \cite{Barnett2000} of the presence or absence of an object given a click or otherwise at a detector. In Section 6 we describe a simple count probability-matching based means of increasing the advantage obtained by quantum illumination. Section 7 describes the simulation of the repeated application of the process described in earlier sections to give the multi-shot scenario. Monte-Carlo simulations clearly show the available advantages of quantum illumination with click detection heralding. Finally we add some concluding remarks in Section 8.

\section{Bosonic Gaussian states}
The calculations, both analytic and computational, in this paper were performed in the Gaussian quantum optical formalism. As stated earlier, this can provide a significant computational advantage for the large Hilbert spaces associated with multi-mode quantum optics. Hence we provide a simple bare overview of the relevant formulae here, but direct the reader to the references for detail \cite{Braunstein2005, Ferraro2005, Weedbrook2012}. In the later sections, however, we typically also provide, for convenience, the standard quantum optical formulae in terms of operators.

For $n$ quantum harmonic oscillator modes, in natural units with $\hbar=1$, a global quadrature operator can be written in the following alternating order
\begin{equation}
	\hat{\text{x}} = \{\hat{q}_1, \hat{p}_1,\ldots,\hat{q}_n,\hat{p}_n\}^\trans,
\end{equation}
where $\hat{q}_k = \frac{1}{\sqrt{2}}(\hat{a}_k+\hat{a}_k^\dagger)$ and  $\hat{p}_k = \frac{i}{\sqrt{2}}(\hat{a}_k-\hat{a}_k^\dagger)$ are the position and momentum quadrature operators of $k$-th mode ($1 \leq k \leq n$), that are non-commuting: $[\hat{q},\hat{p}] = i$. The global commutation relation is
\begin{equation}
	[\hat{\text{x}}_k, \hat{\text{x}}_l] = i \Omega_{kl},\qquad \Omega = \bigoplus_n \begin{pmatrix}
	0 & 1\\
	-1 & 0
	\end{pmatrix},
\end{equation}
where $\Omega$ is the global symplectic identity (symplectic form), that shows the basic phase space structure. The density matrix of a state $\hat{\rho}$ can be mapped to the phase-space Wigner function via a $2n$-dimensional Fourier transform
$\mathcal{W}_{\hat{\rho}}(\text{x}) := \mathcal{F}\left\{\mathcal{X}_{\hat{\rho}}(\Lambda)\right\}$
where $\mathcal{X}_{\hat{\rho}}(\Lambda) := \langle\hat{D}(\Lambda)\rangle$ is the symmetrically-ordered characteristic function of the state \cite{Barnett2003}, that is, the average of the displacement operator $\hat{D}(\Lambda)$, with $\Lambda = \tfrac{1}{\sqrt{2}}\{\Re(\xi_1), \Im(\xi_1),\ldots, \Re(\xi_n), \Im(\xi_n)\}$ the vector of real-numbers corresponding to complex global displacement $\xi = \{\xi_1,\xi^*_1,\ldots,\xi_n,\xi^*_n\}$. If the Wigner function is a Gaussian distribution, then the characteristic function must also be Gaussian. Corresponding states are therefore called Gaussian states $\hat{\rho}_G$ and they are completely characterized by the first and second statistical moments of the Wigner functions. These are expressed by the mean $\mu := \langle\hat{\text{x}}\rangle$ and the covariance matrix $\Sigma$, which has entries
\begin{equation}
	\Sigma_{kl} := \tfrac{1}{2}\langle \hat{\text{x}}_k \hat{\text{x}}_l + \hat{\text{x}}_l \hat{\text{x}}_k \rangle - \mu_k\mu_l.
\end{equation}
For a single mode Gaussian state, $n=1$, in this phase-space picture, $\mu$ denotes displacement of the state from the vacuum (phase space origin) and the covariance matrix describes the widths of the distribution, the uncertainty surrounding $\mu$. All proper Gaussian states exhibit quadrature noise when viewed in the Wigner function representation. In relation to physical parameters, an isotropic spread of the covariance above that of the vacuum state the  indicates the presence of extra thermal noise. Unequal distribution of the Gaussian widths along perpendicular axes indicates noise squeezing, for which the quadrature uncertainties must obey Heisenberg's uncertainty relation $(\Delta \hat{q})^2 (\Delta \hat{p})^2 \geq 1/4$.

Conveniently the Gaussian state's Wigner function exists in non-integral form, as a multivariate Gaussian
\begin{equation}
	\label{eq:wigfunnoint}
	\mathcal{W}_{\hat{\rho}_G}(\text{x}) = \frac{\exp{-\frac{1}{2}(\text{x}-\mu)^\trans\Sigma^{-1}(\text{x}-\mu)}}{(2\pi)^n\sqrt{\det\Sigma}}.
\end{equation}
State overlaps become convolution integrals in phase-space
\begin{equation}
	\tr(\hat{\rho}\hat{\sigma}) = (2\pi)^n\int_{\mathbb{R}^{2n}}d^{2n}\textsc{x}\;\mathcal{W}_{\hat{\rho}}(\textsc{x}) \mathcal{W}_{\hat{\sigma}}(\textsc{x}). \label{eq:wigint}
\end{equation}
Evolution of the quadrature operators via Gaussian unitary operations, that is, unitaries with creation and annihilation operators up to the second order in the interaction Hamiltonian, evolve via $\hat{\text{x}}_k^\prime \rightarrow\hat{U}\hat{\text{x}}_k\hat{U}^\dagger$. These interactions become symplectic transforms on $\mu$ and $\Sigma$
\begin{subequations}
	\begin{eqnarray}
	&\mu^\prime \rightarrow \text{S}^\trans\mu + \mu_D,\\
	&\Sigma^\prime \rightarrow \text{S}\Sigma\text{S}^\trans,
	\end{eqnarray}
\end{subequations}
where $\text{S}$ is the symplectic matrix (a real $2n\times2n$ matrix satisfying $\text{S}\Omega \text{S}^\trans = \Omega$ and $\det\text{S} = 1$) that represents $\hat{U}$ in phase space. The vector $\mu_D$ is a translation caused by the displacement operation. The covariance $\Sigma$ is symplectically diagonalized by casting it in the form  $\text{S}\Sigma^\oplus \text{S}^\trans$ such that $\Sigma^\oplus$ is a diagonal matrix with repeated eigenvalues $\nu_k = \bar{n}_k +\frac{1}{2}$,  $\nbar_k=\langle\hat{a}^\dagger_k\hat{a}_k\rangle$, that shows the global Gaussian state comprising of irreducible thermal states. Tensor products of modes in Gaussian formalism equate to direct sums of moments and to effect partial tracing one can simply delete moments of the traced mode.

If a state is sent through an attenuating channel that also injects thermal noise it undergoes the transformation
\begin{equation}
	\label{eq:losschannel}
	\hat{\rho}^\prime  = \tr_\textsc{e}\left(\hat{U}_\eta\hat{\rho}\otimes\hat{\rho}_{th}\hat{U}_\eta\right),
\end{equation}
where 
\begin{equation}
\label{thermalinput}
\hat{\rho}_{th}(\bar{m}) = \frac{1}{1+\bar{m}}\sum_{n=0}^\infty \left( \frac{\bar{m}}{1+\bar{m}} \right)^n |n \rangle \langle n|
\end{equation}
is a thermal state with average photon number $\bar{m} = \nbar_{th}/(1-\eta)$ and the mixing of two modes is performed by $\hat{U}_\eta = \exp{i\theta(\hat{a}^\dagger\hat{a}_{th}+{\hat{a}^\dagger}_{th}\hat{a})}$, which is a two-mode rotation operator with mixing angle $\theta = \cos^{-1}\sqrt{\eta}$ that also models the beamsplitter with intensity transmissivity $\eta$. The factor $(1-\eta)^{-1}$ in Eq. \eqref{thermalinput} is required to scale the mean photon number of the thermal state such that the output mode of the beamsplitter, traced over the environment, appears to be in a thermal state of mean photon number $\nbar_{th}$ when the channel input state is the vacuum state. For the Gaussian state that has undergone loss $\hat{\rho}^\prime _G$, we simply need to transform its moments $\mu$ and $\Sigma$
\begin{subequations}
	\label{eq:covlossmap}
	\begin{eqnarray}
	&\mu^\prime = \tr_\textsc{e} \left(\text{S}^\trans_\eta(\mu\oplus 0)\right), \label{seq:meanloss}\\
	&\Sigma^\prime  = \tr_\textsc{e} \left(\text{S}_\eta(\Sigma\oplus\Sigma_{th})\text{S}^T_\eta\right), \label{seq:covloss}
	\end{eqnarray}
\end{subequations}
where $\tr_\textsc{e}$ is the partial trace over the environment noise mode (which here denotes simply removing appropriate rows and columns), $\text{S}_\eta$ as the symplectic beamsplitter transform and $\Sigma_{th}$ the covariance of a thermal mode with average photon number $\nbar_{th}/(1-\eta)$, which has zero mean $\mu_{th} = 0$.

As a simple example illustrating the above, a single mode thermal state with mean photon number $\nbar$ (c.f.$\,$ Eq. \eqref{thermalinput}) has Gaussian Wigner function $\mathcal{W}(\text{x}) = \frac{2}{\pi(1+2\nbar)}\exp{-\frac{|\text{x}|^2}{2\nbar+1}}$ and covariance matrix $\Sigma = (\nbar+\frac{1}{2})\mathbbm{1}_2$. When sent through an attenuating channel that contains thermal noise of mean photon number $\nbar_{th}/(1-\eta)$, the covariance matrix becomes
\begin{equation}
    \Sigma^\prime  = \left(\eta\nbar + \nbar_{th} +\frac{1}{2}\right)\mathbbm{1}_2,
\end{equation}
corresponding to a Wigner function of
\begin{equation}
    \mathcal{W}(\text{x}) = \frac{2}{\pi(1+2\eta\nbar + 2\nbar_{th})}\exp{-\frac{|\text{x}|^2}{1+2\eta\nbar + 2\nbar_{th}}},
\end{equation}
that is, a broadened Gaussian if $\nbar_{th} > \nbar(1-\eta)$. In addition, the original state has its average photon number attenuated by a factor $\eta$.

\section{Direct photodetection}
Our direct photodetection measurement consists of just two outcomes: click or no-click, in other words the detector fires or it does not. Usually in quantum information, the measurement outcome is the expectation value of a POVM operator, so that, for a state $\hat{\rho}$ measured by an imperfect click detector with a dark count probability (the probability that the detector fires if no light falls upon it) provided by a thermal distribution with $\nbar_d > 0$ and quantum efficiency $\eta < 1$ \cite{Rohde2006},
\begin{equation}
	\Pr(\times|\hat{\rho}) = \tr\left(\hat{\Pi}_\times(\eta, \nbar_d)\hat{\rho}\right)
\end{equation}
is the no-click probability. The operator $\hat{\Pi}_\times$ is the no-click operator
\begin{equation}
    \label{eq:noclickpovm}
    \hat{\Pi}_\times(\eta,\nbar_d) = \frac{1}{1+\nbar_d}\sum_{n=0}^\infty \left(1-\frac{\eta}{1+\nbar_d}\right)^n\ketbra{n},
\end{equation}
and $\hat{\Pi}_\checkmark = \mathbbm{1} - \hat{\Pi}_\times$ is the click operator corresponding to a dark count click probability with no light incident on the detector of $\nbar_d /(1+\nbar_d)$. By this construction, all click probabilities can be defined in terms of their complementary no-click probabilities. Furthermore, giving the dark count probability distribution a thermal Gaussian character is the essential property that allows the analysis to be performed simply in the Gaussian framework. Dark counts occur randomly and are sometimes modelled via Poisson statistics \cite{Loudon2000,Barnett1998}, but all valid models amount to measurement via projection on to a mixed state. The detailed model used can be adjusted to give the correct no-click probability of the actual detector for no incident light. This adjustment would result in slightly different probabilities for higher numbers of clicks, depending on the model used. Geiger-mode detectors can click once per shot only, no matter how many photons are incident, so only for photon number resolving detectors would the detailed probability distribution of dark counts be noticeable.

Perfect click measurement occurs when $\nbar_d = 0$ and $\eta = 1$, with vacuum projector
\begin{equation}
	\hat{\Pi}_\times(1, 0) = \ketbra{0}.
\end{equation}
The vacuum state is a Gaussian state with covariance matrix $\Sigma=\tfrac{1}{2}\mathbbm{1}$, so the optimal no-click detection of a Gaussian state $\hat{\rho}_G$ depends simply on the overlap between the state and the vacuum,
\begin{equation}
	\label{eq:noclickprob}
	\Pr(\times|\hat{\rho}_G) = {\sqrt{\det(\Sigma+\tfrac{1}{2}\mathbbm{1})^{-1}}}\exp{-\frac{1}{2}\mu^\trans\left(\Sigma+\frac{1}{2}\mathbbm{1}\right)^{-1}\mu},
\end{equation}
derived using Eqns. \eqref{eq:wigfunnoint} and \eqref{eq:wigint}. As the detection operator is the vacuum projector this probability is the Husimi Q-function of a Gaussian state at the phase space origin. For the more general result that includes nonzero dark noise and non-unit quantum efficiency --  instead of converting Eq. \eqref{eq:noclickpovm} into a Wigner function we simply decompose measurement of the state by the \emph{imperfect} click detector into pre-attenuation of the state measured by a \emph{perfect} click detector, as these provide the same outcome
\begin{equation}
	\tr(\hat{\Pi}_\times(\eta, \nbar_d)\hat{\rho}_G) = \tr(\hat{\Pi}_\times(1, 0)\hat{\rho}_G^\prime ).
\end{equation}
As before, the prime indicates $\hat{\rho}_G$ has passed through the loss channel in Eq. \eqref{eq:losschannel} that is easier to implement using the continuous-variable formalism presented in Eqns. \eqref{eq:covlossmap}.


\section{Conditioned single mode states}
The TMSV is an entangled Gaussian state which exhibits photon number and quadrature correlations (with zero-mean field). It has the following Fock basis wavefunction \cite{Barnett1987}
\begin{equation}
	\ket{\Psi}_\textsc{i,\,s} = \sqrt{1-\lambda^2}\sum_k \lambda^k \ket{k,k},
\label{twomode}
\end{equation}
containing modes $\textsc{i}$ - idler and $\textsc{s}$ - signal corresponding to different beams of light propagating, in principle, in different directions. The factor $\lambda = \sqrt{\frac{\nbar}{1+\nbar}}$ contains $r$ as the squeezing amplitude in the form of the single mode average photon number $\nbar := \sinh^2r$. Each beam or arm of the TMSV contains an average $\nbar$ photons.

\begin{figure}[t!]
	\centering
	\includegraphics[width=0.7\linewidth]{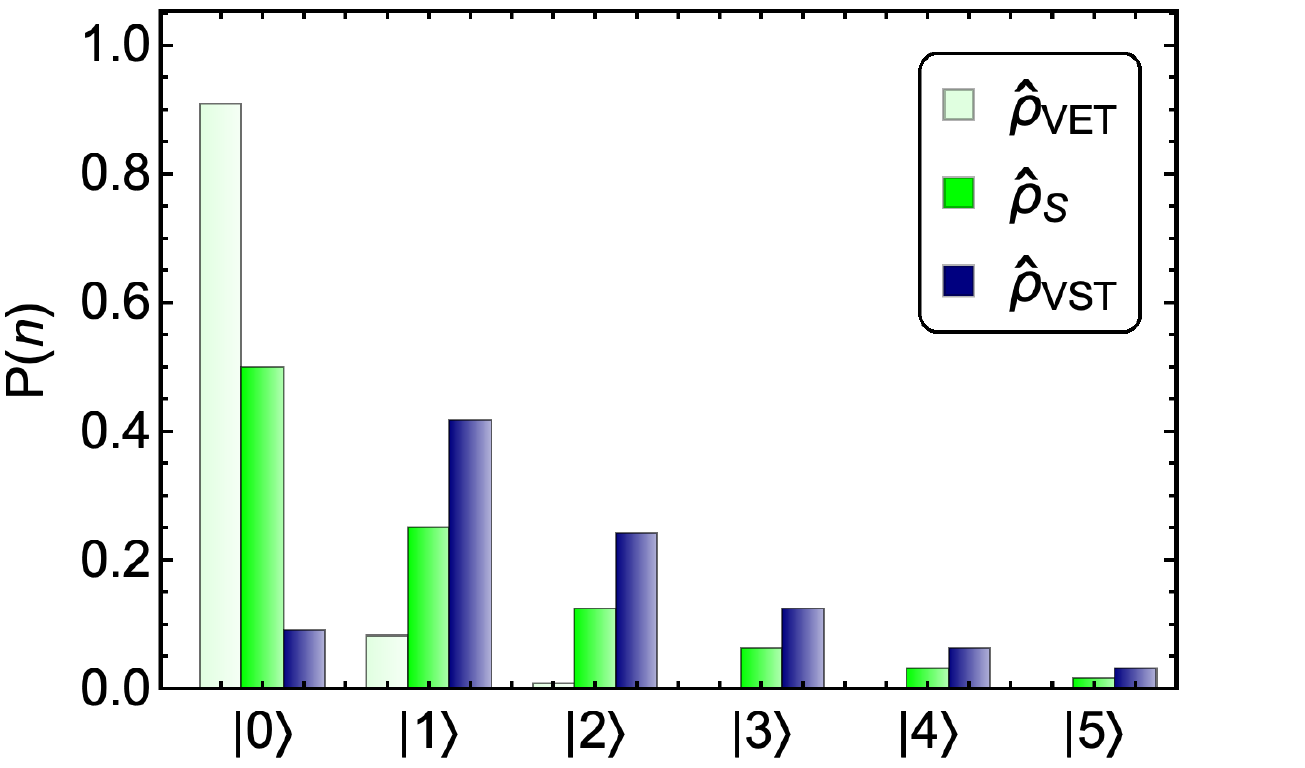}
	\caption{Photon number distribution of conditioned single mode states: $\nbar = 1$, $\nbar_{d_\textsc{i}} = 0.9$ and $\eta_\textsc{i} = 0.1$ are the dark-count and efficiency of the idler detector. Vacuum suppression is evident here as dark counts are low and efficiency is high. The photon number suppressed state is also shown. If a click detector were to perform a subsequent measurement on this state, then its results would be correlated with the heralding detector.}
	\label{fig:PhotonDists}
\end{figure}

The two outcomes of click detection on the idler mode of the TMSV herald two different conditional signal states. When the no-click outcome occurs, the remaining signal state is conditioned into
\begin{equation}
	\hat{\rho}_\textsc{pnst} = \mathcal{N}^{-1}\tr_\textsc{i}\big(\hat{\Pi}_{\times \textsc{i}}(\eta, \nbar_d) \hat{\rho}_\textsc{i,\,s}\big),
\end{equation}
where PNST denotes the photon number suppressed thermal state and $\hat{\rho}_{\textsc{i,\,s}}= \ketbra{\Psi}_\textsc{i,\,s}$. The normalisation is simply the no-click probability from Eq. \eqref{eq:noclickprob} of the idler mode: $\mathcal{N}=\Pr(\times|\hat{\rho}_\textsc{I}^\prime )$, which may contain terms relating to the dark-noise and quantum efficiency depending on the optimality of the heralding measurement. 

When a click occurs, the remaining signal state is conditioned to become
\begin{equation}
	\hat{\rho}_{\textsc{vst}} = (1-\mathcal{N})^{-1}(\hat{\rho}_\textsc{s} - \mathcal{N} \hat{\rho}_{\textsc{pnst}}),\label{eq:vstwavefunc}
\end{equation}
which is the VST - vacuum suppressed thermal state. 
The PNST state is a thermal state with mean photon number lower than $\bar{n}$, the unconditioned mean. It has statistics similar to the partially traced TMSV $\hat{\rho}_\textsc{s} = \tr_\textsc{i}\hat{\rho}_{\textsc{i,\,s}}$ (see Fig. \ref{fig:PhotonDists}), which is a thermal state with covariance matrix $\Sigma_\textsc{s}=(\nbar+\tfrac{1}{2})\mathbbm{1}$. The VST state is not thermal, but it can be expressed by Eq. \eqref{eq:vstwavefunc} as a weighted difference of the unconditioned thermal state and the PNST state. 

The photon number distributions of the average and the conditioned states are shown in Fig. \ref{fig:PhotonDists}. It is clear from this figure that the no click result conditions the signal to be in a state with a smaller mean photon number, the PNST. Consequently the click result conditions the signal to be in a state with a higher mean photon number, the VST. The postselection-based conditioning of the state is a direct consequence of the correlations between the signal and idler beams. The correlations are quantum in nature: the variance in the difference in photon numbers between the beams is zero for the state given by Eq. (\ref{twomode}). Therefore a click result, which measures the idler beam in a mixed state containing at least one photon, forces the signal beam into a state with increased mean photon number.The relative size of this increase becomes larger as $\nbar \rightarrow 0$. It is this increase in photon number that we can exploit for object detection. If optimal heralding detection is performed, with a perfect photodetector, the two conditioned signal states become $\hat{\rho}_\textsc{pnst} \rightarrow \ketbra{0}$ and $\hat{\rho}_\textsc{vst} \rightarrow \frac{1}{\nbar}\left(\sum_k (\frac{\nbar}{1+\nbar})^k\ketbra{k}-\ketbra{0}\right)$ -- the latter state is completely vacuum removed. In such an ideal case, the vacuum removal increases the average photon number of the remaining signal mode by 1. Any attempt to mimic the effect with classical states will not be as successful. Classically-correlated twin beams can be produced with thermal light and a beamsplitter. Detection of light in one output beam conditions the other beam to have a higher mean photon number, which is the principle behind ghost imaging with thermal light \cite{Padgett2017}. However, Geiger-mode detection typically only increases the mean photon number by a small amount in the low mean photon number regime. 
An attempt to produce the same effect with a coherent state input to a beamsplitter will not work as the two-mode output is factorizable.

\begin{figure}[t]
	\centering
	\includegraphics[width=0.8\linewidth]{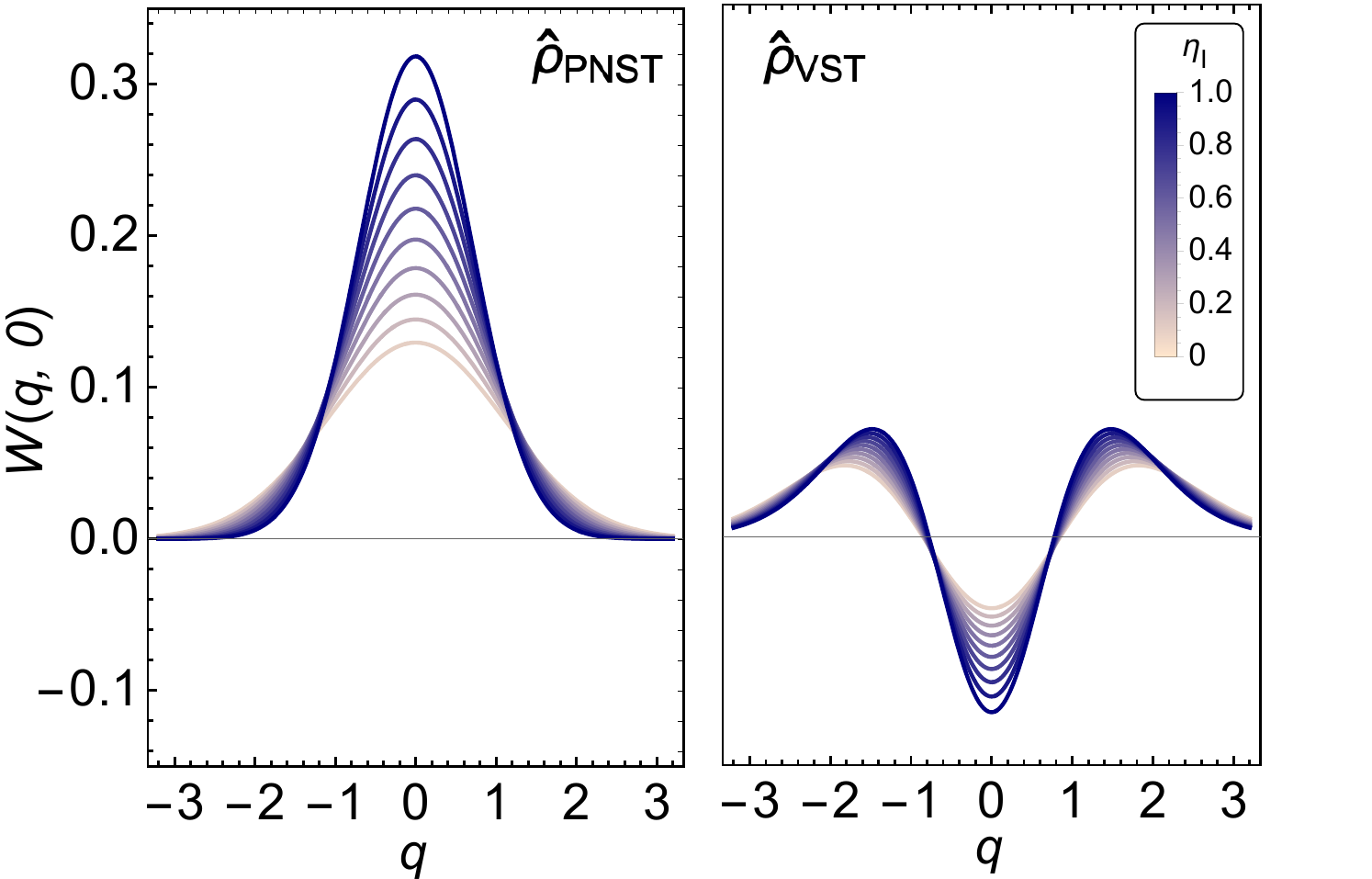}
	\caption{Slices through Wigner functions of $\hat{\rho}_\textsc{pnst}$ and $\hat{\rho}_\textsc{vst}$ showing the effect of varying heralding efficiency (without dark noise). Imperfect heralding efficiency will still cause vacuum suppression. Negativity of the Wigner function is exhibited only by non-classical states \cite{Kenfack2004}.}
	\label{fig:wigpnstvst}
\end{figure}

The Wigner functions of the conditional states can be found from the covariance matrix of the TMSV,
\begin{equation}
	\label{eq:tmsvcov}
	\Sigma_\textsc{i,\,s} = \begin{pmatrix}
	V_\textsc{i} & C\\
	C & V_\textsc{s}
	\end{pmatrix},
\end{equation}
that has sub-matrices $V_\textsc{i,\,s} = \left(\nbar+\tfrac{1}{2}\right)\mathbbm{1}_2$ and $C = \sqrt{\nbar(1+\nbar)}\mathbbm{1}_2$ for its quadrature correlations. This hollow covariance matrix is in a so-called standard form and can always be achieved from an  arbitrary covariance matrix through a sequence of local or global rotation and squeezing symplectic transformations \cite{Duan2000}. By performing a partial trace integral in phase space, we can extract the covariance matrix of the PNST as
\begin{equation}
	\Sigma_{\textsc{pnst}} = V_\textsc{s}-C^T\left(V_\textsc{i}+\tfrac{1}{2}\mathbbm{1}_2\right)^{-1}C,\label{eq:noclickcov}
\end{equation}
which resembles a Schur complement matrix, indicating also that the PNST is Gaussian. All of the states that we require can be represented as Gaussian thermal states or weighted mixtures of them. This allows calculations to be performed with thermal states in the Gaussian quantum optical formalism, as was stated in the Introduction, even though the VST state is not Gaussian, as shown by the Wigner functions in Fig. \ref{fig:wigpnstvst}. If imperfect heralding occurs, then we can apply losses to the idler covariance $V_\textsc{i}\rightarrow V^\prime _\textsc{i}$ via Eq. \eqref{seq:covloss}. The Wigner function of the non-Gaussian VST is then obtained by transforming the individual density matrices in Eq. \eqref{eq:vstwavefunc} into Wigner functions by using $V_\textsc{s}$ and $\Sigma_{\textsc{pnst}}$. This provides a weighted difference of Wigner functions
\begin{equation}
	\label{eq:wigfunvst}
	\mathcal{W}_\textsc{vst}(\text{x}) = (1-\mathcal{N})^{-1}\Big(\mathcal{W}_\textsc{s}(\text{x}) - \mathcal{N} \mathcal{W}_\textsc{pnst}(\text{x})\Big),
\end{equation}
weighted by the no-click probability from the idler mode $\mathcal{N}$. This expression is useful for calculating the probabilities in the next section, as well as for demonstrating that non-Gaussian states can sometimes be written as weighted sums of Gaussian states.


\section{Quantum illumination for target detection}
The illumination and detection procedure outlined above amount to the application of quantum hypothesis testing in a scenario where the detection is limited to Geiger-mode click or no-click devices. We need to decide between the two hypotheses, $H_0$ - the target is absent and $H_1$ - the target is present, based on a set of detector clicks. In this section we will analyse the effect on our confidence) in the hypotheses of sending one state, either quantum ($\hat{\rho}_{\textsc{vst}}$) or classical ($\ket{\alpha}$), to investigate the target area. We use the detection result at the reflected signal detector to update our prior probability that the target is present. When a target object is present more light is reflected to the detector so it should be more likely to fire, whether we send classical light or quantum. We will assume that each hypothesis is equally-likely {\it a priori}. In Section 7 we will show the effect of repeated application of this procedure, so that the outcome of one experiment becomes the nonuniform prior for the next. 

Our procedure is a detection-limited form of the hypothesis testing or state discrimination problems, which have traditionally been formulated in terms of error probabilities. Here we calculate the conditional probabilities of detection events given the presence or absence of the target for either a quantum or classical probe. We then use these and the prior to provide retrodictive conditional probabilities of the two hypotheses, given either a click or no-click result at the detector. This separates the two kinds of errors: the probability that the target is not present given that the detector fires and the probability that the target is present given that the detector does not fire. In principle it can allow us to prioritize one over the other.

Consider a scenario where we send a single mode state $\hat{\rho}$ as a probe to sense the possible presence of an object with reflectivity $\kappa$ ($0<\kappa<1$) that is in a thermal noise bath. The probe has an average photon number $\nbar$ and the noise is modelled by the thermal state $\hat{\rho}_b$ with mean thermal photon number $\bar{n}_b$. Conditioned by the presence or absence of the object, we eventually have to discriminate between two different possible return states
\begin{subequations}
	\begin{eqnarray}
	&\hat{\rho}_0 = \hat{\rho}_b,\\
	&\hat{\rho}_1 = \tr_\textsc{E}\left( \hat{U}_\kappa\hat{\rho}\otimes\hat{\rho}^\prime _{b}\hat{U}_\kappa^\dagger\right), \label{seq:h1densmat}
	\end{eqnarray}
\end{subequations}
where $\hat{\rho}_0$ is the thermal background state $\hat{\rho}_{b}$ with $\bar{n}_b$ that our detector receives if the object is absent (the $H_0$ state); $\hat{\rho}_1$ is the state containing reflected signal photons that we receive if the object is present (the $H_1$ state).  Notice that $\hat{\rho}_1$ is formed via the same transformation as Eq. \eqref{eq:losschannel} with $\eta$ replaced by $\kappa$ and that $\hat{\rho}^\prime _{b}$ has scaled average photon number $\nbar_{b}/(1-\kappa)$.

A click detection measurement will satisfy the following: if the object is absent, the detector will fire with what could be called a false alarm probability dependent on the thermal background
\begin{equation}
	\Pr(\checkmark|\hat{\rho}_0) = 1 - \frac{1}{1+\nbar_d+\eta\nbar_b},
\end{equation}
independent of any object properties $\kappa$ or signal photon number $\nbar$. There is no object to detect so the signal does not reach the detector, hence this result applies to all possible sent signal states. The factors $\eta$ and $\nbar_d$ are the receiving detector efficiency and dark noise.

If the object is present, then sending a coherent state signal $\hat{\rho}_\alpha$ that has $\nbar=|\alpha|^2$, $\mu = \{\sqrt{2\bar{n}}, 0\}^T$ and $\Sigma = \tfrac{1}{2}\mathbbm{1}_2$ gives a click probability for $\hat{\rho}_1$
\begin{equation}
	\Pr(\checkmark|\hat{\rho}_1)_{\alpha} = 1 - \frac{1}{1+\bar{n}_d + \eta\nbar_b}\exp{-\frac{\eta\kappa\nbar}{1+\nbar_d+\eta\nbar_b}}.
\end{equation}
If the target object is present and we are instead using the heralded TMSV as our signal state, then the single mode state sent becomes one of the two conditioned states caused by click-heralding of the idler. If our local detector does not fire, then the receiving detector has a click probability of 
\begin{equation}
	\Pr(\checkmark|\hat{\rho}_1)_\textsc{pnst} = 1 - \sqrt{\det(\Sigma^{\prime \prime} _{\textsc{pnst}}+\tfrac{1}{2}\mathbbm{1}_2)^{-1}},
\end{equation}
where the double prime on the covariance indicates the application of two loss channels: once through the object, once again for detector losses. Its explicit expression is
\begin{equation}
	\Sigma^{\prime \prime} _\textsc{pnst} = \left(\frac{1}{2}-\eta\kappa+\eta\nbar_b +\nbar_d + \frac{\eta\kappa(1+\nbar)(1+\nbar_{d_\textsc{i}})}{1+\nbar_{d_\textsc{i}}+\nbar\eta_\textsc{i}}\right)\mathbbm{1}_2,
\end{equation}
which contain the idler detector efficiency $\eta_\textsc{i}$ and dark noise $\nbar_{d_\textsc{i}}$. If the local idler detector fires, then we send the state $\hat{\rho}_\textsc{vst}$, which enhances the click probability of the receiving detector with a probability of 
\begin{equation}
	\Pr(\checkmark|\hat{\rho}_1)_\textsc{vst} = 1 - \frac{1}{1-\mathcal{N}}\left(\sqrt{\det(V_\textsc{s}^{\prime \prime} +\tfrac{1}{2}\mathbbm{1}_2)^{-1}} - \mathcal{N}\sqrt{\det(\Sigma^{\prime \prime} _\textsc{pnst}+\tfrac{1}{2}\mathbbm{1}_2)^{-1}}\right),
\end{equation}
where 
\begin{equation}
	V_\textsc{s}^{\prime \prime}  = \left(\frac{1}{2} + (\nbar_b+\nbar\kappa)\eta+ \nbar_{d} \right)\mathbbm{1}_2,
\end{equation}
is the thermal covariance of the vacuum suppressed state.

\begin{figure}[t!]
	\centering
	\includegraphics[width=0.9\linewidth]{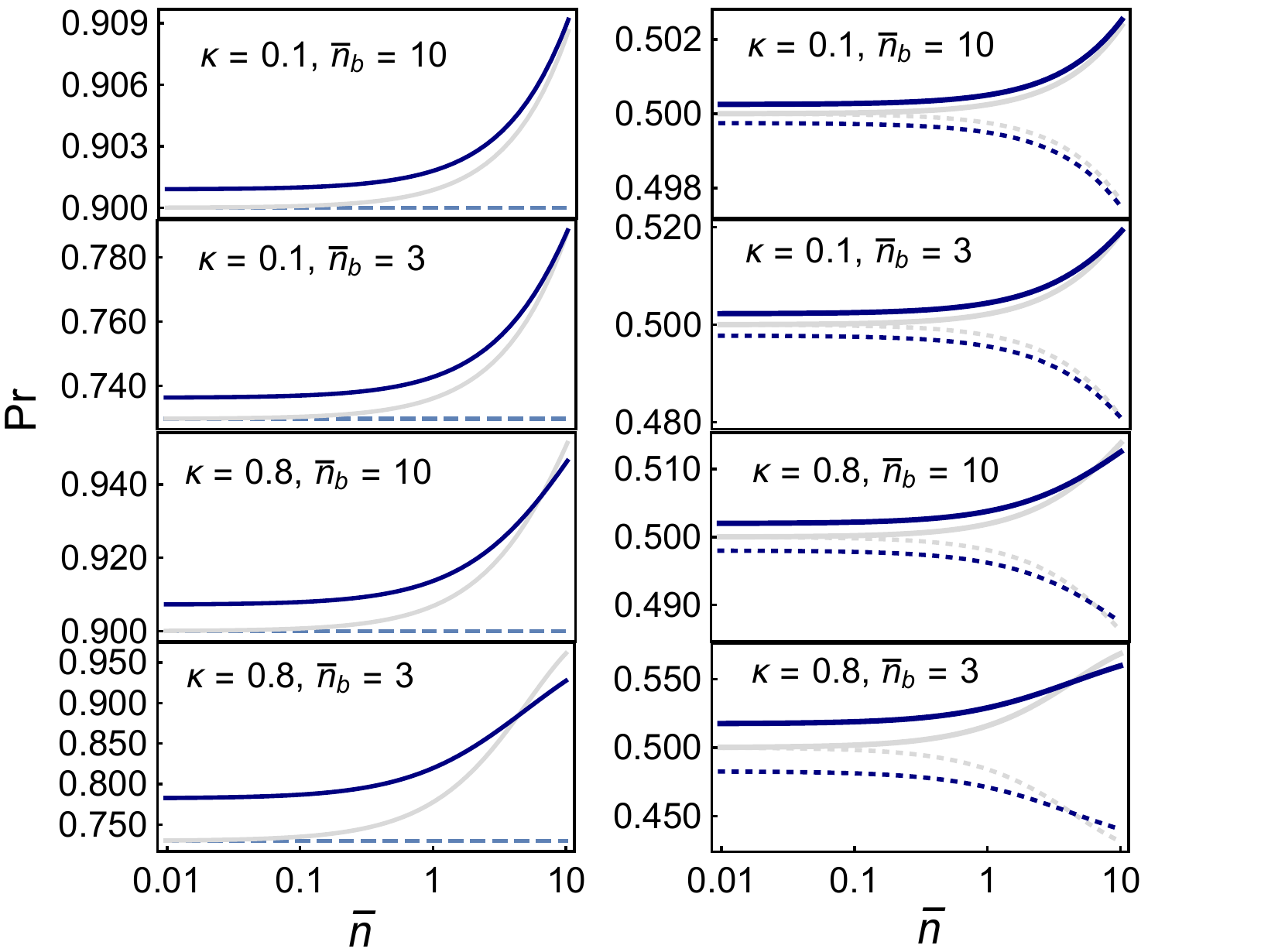}
	\caption{Plots showing click probability of the receiving signal detector ($\nbar_d=0, \eta_\textsc{i}=\eta = 0.9)$ against increasing average signal photon number $\nbar$. Each figure shows varying object reflectivity $\kappa$ and background noise $\nbar_b$. LEFT: Conditional click probability for $\hat{\rho}_0$ (dashed) or $\hat{\rho}_1$ (pink - $\hat{\rho}_\alpha$, navy - $\hat{\rho}_\textsc{vst}$). RIGHT: Posterior probability updated from prior probabilities of $\Pr(H_0)=\Pr(H_1)=1/2$ after a click has occurred. Dotted and solid curves indicate $\Pr(H_0)$ and $\Pr(H_1)$ respectively.}
	\label{fig:clickprobs}
\end{figure}

Plots of click probabilities for a detector receiving $\hat{\rho}_0$ vs. $\hat{\rho}_1$ caused by sending different states can be found in Fig. \ref{fig:clickprobs}. The left column of plots shows the probability of obtaining a count as a function of mean signal photon number. In most cases the background photon number is much higher than that of the signal. The no-object probability $\hat{\rho}_0$ is a flat line as it does not depend on the signal photon number. There is a noticeable heralding gap between the probability obtained using a coherent signal state and the vacuum suppressed state derived from a heralded TMSV of the same mean photon number. The advantage is largest for low $\bar{n}$ as heralding has a greater effect, suggesting that quantum illumination is advantageous for low average signal photon number. The effect persists even for higher loss scenarios, but in all cases the gap closes as $\nbar$ increases and eventually the coherent state outperforms the vacuum suppressed state. The reason for this is that the heralding has little effect on the thermal distribution for high $\nbar$ and the Poissonian nature of the coherent state photon number distribution peaks there. This effect dominates over the weak heralding effect. 

The right column of plots shows the probability that we would assign to the presence $H_1$ or absence $H_0$ of an object based on a click at the receiving detector as a function of mean signal photon number. We assume no prior knowledge of the presence of the object. A click at the detector increases the probability that the object is present and decreases the probability that it is absent. Again there is an gap at low mean photon number between heralded TMSV and coherent light, showing the quantum illumination advantage. Again the coherent state wins out as the mean photon number of the sent signal state increases.

\begin{figure}[t!]
    \centering
    \includegraphics[width=0.7\linewidth]{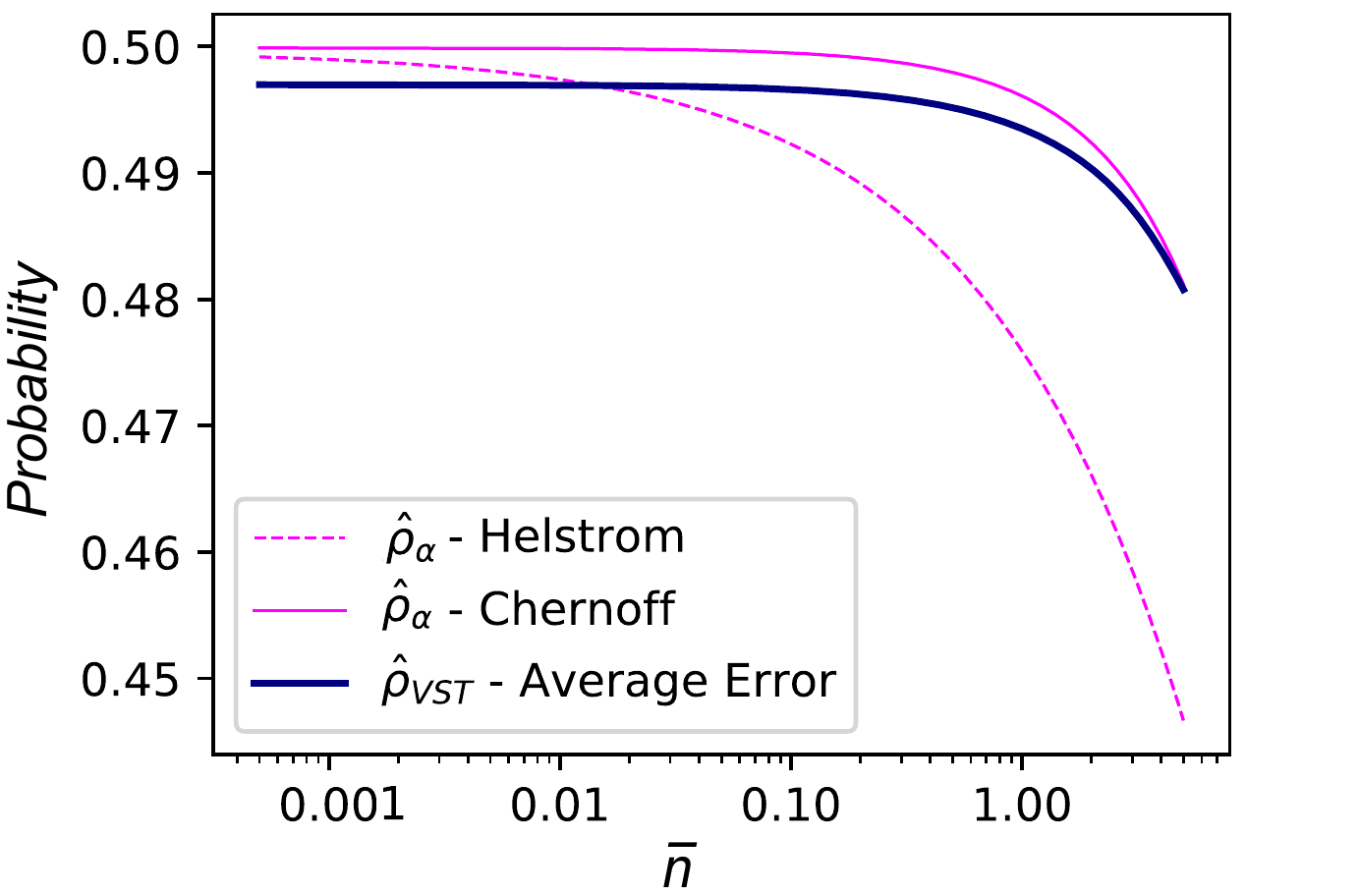}
    \caption{Single-shot error probability for the signal detector: the performance of click-heralded $\hat{\rho}_\textsc{vst}$ state against the coherent state quantum Chernoff bound and Helstrom bound are compared here. Parameters: $\kappa = 0.1$ and $\nbar_b = 3$. The average error from using click-measurement is $\tfrac{1}{2}\Pr(\times|\hat{\rho}_1) + \tfrac{1}{2}\Pr(\checkmark|\hat{\rho}_0)$, and successful heralding leads to a slightly lower error rate below that of coherent state Helstrom bound. Quantum Chernoff bound and Helstrom bound should coincide to 1/2 when $\nbar = 0$.}
    \label{fig:PerrBounds}
\end{figure}

In Fig. \ref{fig:PerrBounds} the relation between the measurement described here and the fundamental Helstrom bound is shown. We plot the average error probability in distinguishing the click heralded state after it has been reflected from a target from the thermal background state. This is compared with the Helstrom bound for distinguishing a coherent state reflected from the target and the thermal background state. For completeness we also show the higher Chernoff bound, which is sometimes used for ease of calculation. At very low mean photon numbers the click-heralded state beats the Helstrom bound, but the fact that the click-measurement is not ideal means that for higher photon numbers this advantage is lost, as well as the heralding probability for TMSV at very low $\nbar$ is negligible. Of course the Helstrom measurement is likely to be impractical, for example in a rangefinding lidar scenario. Then, the appropriate comparison is given by Fig. \ref{fig:clickprobs}.


\section{Click probability matching}
The previous section shows clearly the single-shot advantage of quantum illumination for object detection with click detection, but only allowing click detection confers other advantages.  We can also exploit the fact that the average signal photon number distribution has particular forms, with specific detector click probabilities. These specific forms of distribution allow the coherent state to defeat quantum illumination for higher mean signal photon numbers in Fig. \ref{fig:clickprobs}, but we can turn them to our advantage. Rather than comparing the results obtainable with classical coherent and heralded TMSV states of the same mean photon number, we can compare sent states that would give the same click probability at a detector.

\begin{figure}[t!]
	\centering
	\includegraphics[width=\linewidth]{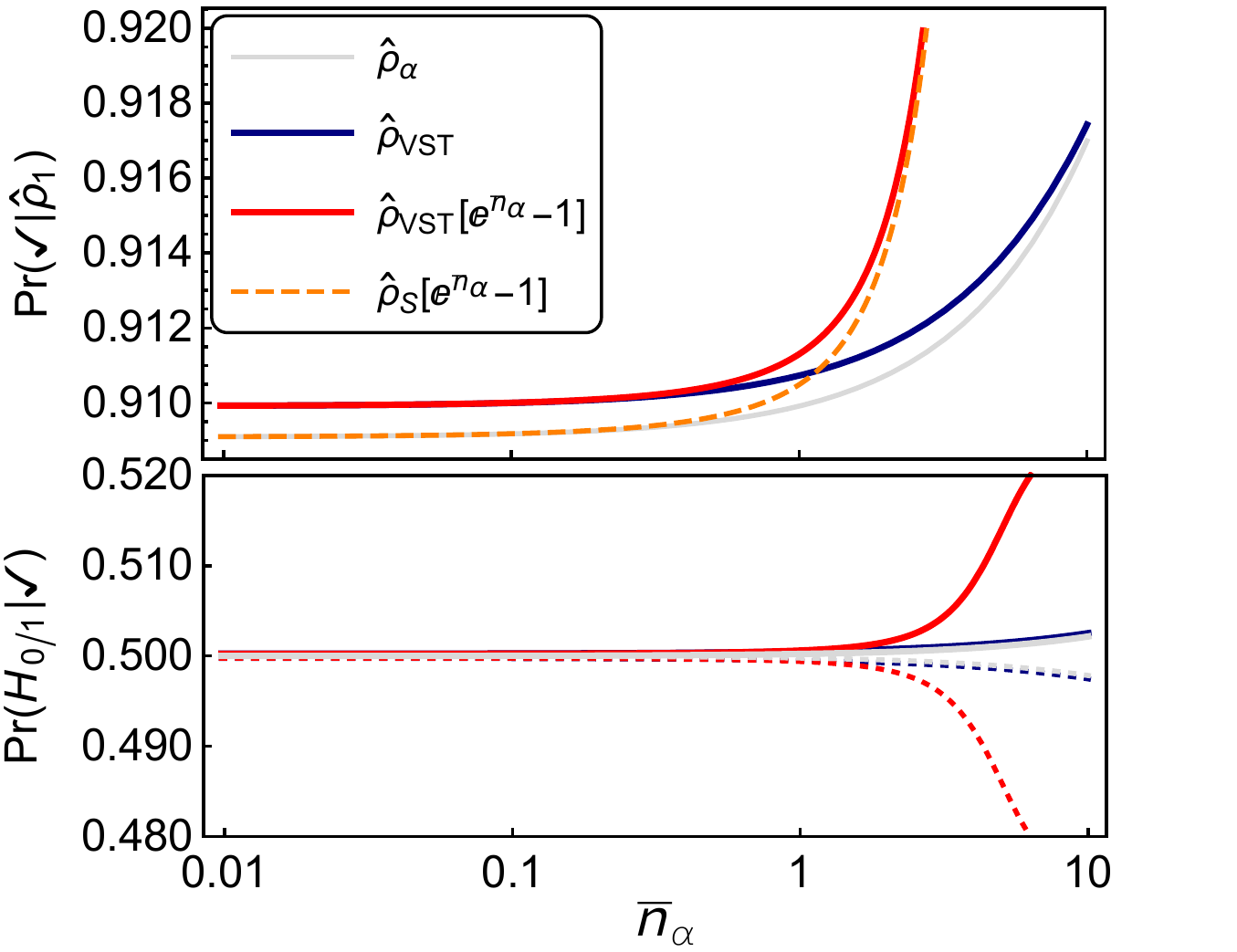}
	\caption{LEFT: Click probabilities showing enhancement obtained by matching the perfect detector click probabilities of the TMSV and the coherent state. RIGHT: Posterior probability update from equal prior probabilities. Also, $\kappa = 0.1$ and $\nbar_{b} = 10$. Both thermal state $\hat{\rho}_\textsc{s}$ and vacuum suppressed curves with average photon number $\exp{\nbar_\alpha}-1$, $\nbar_\alpha = |\alpha|^2$ are more likely to trigger clicks past $\nbar_\alpha=1$, which would allow us to choose a higher energy TMSV without compromising discoverability of our quantum source.}
	\label{fig:MatchedProbs}
\end{figure}

Our justification for this strategy is that with a single Geiger-mode detector the coherent and average quantum states of the same mean photon number are distinguishable from each other (and therefore differently distinguishable from a thermal background). We can render them indistinguishable by matching click probabilities. Such states are then effectively indistinguishable by single detector clicks at Geiger-mode detectors, whatever their specific photon probability distributions. The matching of click probabilities will not compromise source discoverability should we wish to illuminate covertly, rather it will enhance the covertness. In quantum illumination we should compare the ability of states with the same resource to discriminate the target from the background state. Typically in optics this resource is mean photon number, but for discrimination based solely on Geiger-mode detection the relevant resource is effectively click probability.

When one mode of the TMSV is observed independently of the other, it appears to be in a thermal state with average photon number $\nbar_\textsc{s} = \sinh^2r$. If a click detector wishes to intercept the signal before interaction with the object it will fire with probability
\begin{equation}
	\Pr(\checkmark|\hat{\rho}_\textsc{s}) = \frac{\eta\nbar_s}{1+\eta\nbar_\textsc{s}},
\end{equation}
which is lower than the coherent state probability with average photon number $\nbar_\alpha = |\alpha|^2=\nbar_s$
\begin{equation}
	\Pr(\checkmark|\hat{\rho}_\alpha) = 1-\exp{-\eta\nbar_\alpha}.
\end{equation}
Instead of the above choice we can choose $\nbar_\textsc{s}$ to match these click probabilities by equating the above two expressions
\begin{equation}
	\eta\nbar_\textsc{s} = \exp{\eta\nbar_\alpha} - 1,
\end{equation}
which means that
\begin{equation}
	\nbar_\textsc{s} = \nbar_\alpha\left(1 + \frac{\eta\nbar_\alpha}{2!} + \frac{\eta^2\nbar_\alpha^2}{3!} + \ldots\right).
\end{equation}
We are able to increase the mean photon number of the quantum illumination beyond that of the coherent state, without compromising source discoverability by a click detector (see Fig. \ref{fig:MatchedProbs}). Whilst this is not strictly a quantum advantage it does more than offset the classical advantage that the coherent signal state has at higher signal photon numbers, as can be seen by comparing the orange curves and the pink curves in both the left panel (click probabilities) and right panel (object presence or absence probabilities) of Fig. \ref{fig:MatchedProbs}. The heralding advantage dominates at low mean photon numbers, where the potential for click probability matching advantage is limited (the difference between the green dot-dashed and orange curves in the left panel, or the pink and navy in both panels). At signal photon numbers around one the click probability matching advantage begins to dominate. While this is still in the quantum regime it is clear that the click-probability matching provides most advantage at higher mean photon numbers, so it will not be so useful for the lowest mean photon number TMSV quantum lidar systems, nor for detecting systems sensitive to higher energies.

One objection to click probability matching might be that we are limiting the means of discoverability to clicks made at a single detector. Multiple coincident detector clicks would allow the two different sources to be discriminated so, of course, the objection is correct. However, this objection applies to distinguishing the coherent state from {\it any} state that does not have a Poissonian photon number probability distribution. Here we are specifically considering limited click detection as it is the simplest and most likely form in optics. Moreover, the probability of multiple coincident clicks in a real object identification system operating near the quantum level will be tiny. For such systems click probability matching can only decrease the chances of discoverability.

\section{Modelling a sequential detection process}
\label{sequence}
As the small changes in the posterior probabilities from the no-information values of 1/2 in Figs. \ref{fig:clickprobs} and \ref{fig:MatchedProbs} show, single shot experiments provide only a tiny amount of information about the presence of the target. In order to achieve greater confidence in estimation of the object's presence (or absence) we can apply the above process repeatedly in a multi-shot scenario. We herald multiple sequential TMSV states and send them to interact with a possible target before repeated click detection. Experimentally this can be realized by sending a train of light pulses to probe the region of interest. As shown in Fig. \ref{fig:clickprobs}, after each detector result given the object presence (absence), the posterior probability is updated from prior probabilities.  We are able to simulate a repeated update of the estimate of the probability $\Pr(H_1)$ that an object is present, based on click or no-click measurement outcomes at both the idler and measurement detectors. The overall process is simple to simulate numerically and requires only Bayes' Law and a (pseudo)random number generator. 

\begin{figure}
    \centering
    \includegraphics[width=\linewidth]{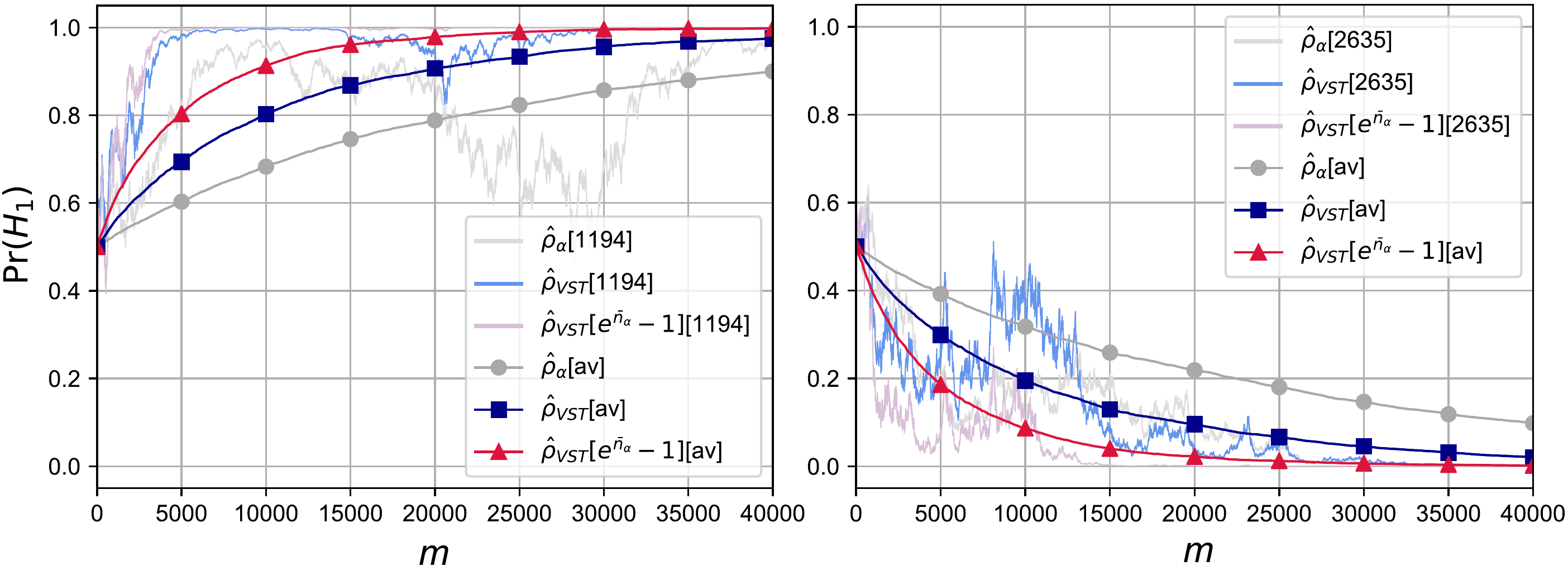}
	\caption{Monte-Carlo trajectories as a function of shot number $m$, from the described procedure in Section \ref{sequence}. Parameters: $\nbar = 1, \eta = \eta_\textsc{i}=0.9, \nbar_b = 3$ and $\kappa = 0.1$. For the $\hat{\rho}_{\textsc{vst}}$ trajectories we take into account of the heralding process. LEFT: object present. RIGHT: object absent. The smooth curves are averages of 3000 different trajectories. Individual trajectories (randomly selected) from the sets are also presented - no.512 (left) and no.1665 (right). Estimated object presence or absence converges correctly to certainty in fewer measurements when quantum illumination is used with click detectors, compared to coherent state illumination.}
	\label{fig:trajectories}
\end{figure}

Suppose that we send a set of sequential states (shots) to the possible target object, making $M$ sequential measurements. We begin by assuming that the prior probability of object presence is $\Pr(H_1) = 1/2$, which can be considered as the zeroth trial $\Pr^{(m=0)}(H_1)$. We want to simulate the entire detection process either when a target object is present or when one is not. First assume that an object is present. In order to model the outcome of the single click-detector iteratively, we throw a uniformly-distributed random number $p$ between 0 and 1 for each $m$-th measurement ($0\leq m\leq M$) and subject it to the following update rule
\begin{equation}
	\text{Pr}^{(m+1)}(H_1) = \begin{cases}
		\text{Pr}^{(m)}(\hat{\rho}_1|\checkmark) & p < \Pr(\checkmark|\hat{\rho}_1)\\
		\text{Pr}^{(m)}(\hat{\rho}_0|\checkmark) & p \geq \Pr(\checkmark|\hat{\rho}_1)
		\end{cases}.
\end{equation}
Essentially, if the random number is smaller than the click probability we infer that the detector has fired. If the random number is larger than the click probability we infer that the detector has not fired. We update the probabilities of object presence or absence accordingly. The updated estimated probability for object absence, which complements that above, is
\begin{equation}
	\text{Pr}^{(m+1)}(H_0) = \begin{cases}
		\text{Pr}^{(m)}(\hat{\rho}_1|\times) & p < \Pr(\checkmark|\hat{\rho}_1)\\
		\text{Pr}^{(m)}(\hat{\rho}_0|\times) & p \geq \Pr(\checkmark|\hat{\rho}_1)
	\end{cases},
\end{equation}
where $\Pr(\hat{\rho}_1|\checkmark)$ is the posterior probability for the state and detection outcomes calculated via Bayes' Law. The above example is with the object is present, as we have used $\Pr(\checkmark|\hat{\rho}_1)$ in the conditional statement. In order to simulate the sequential measurement given that the object is absent, we can switch the probability in the conditional statement to $\Pr(\checkmark|\hat{\rho}_0)$. 

As production of the vacuum suppressed state requires a heralding click detector, we must adapt the procedure by subjecting a second random number, $h$ to the following condition
\begin{equation}
	 h < \Pr(\times|\hat{\rho}_\textsc{i}),
\end{equation}
before moving on to calculate the posterior probabilities such that when no-click occurs on the idler we proceed to calculate conditional probabilities using $\hat{\rho}_{\textsc{pnst}}$. If click heralding occurs we use $\hat{\rho}_{\textsc{VST}}$. 
Eventually, after a number of measurements, the estimated probability will reach convergence, as shown by Fig.  \ref{fig:trajectories}, after which we can conclude that the object is present or absent. We have chosen this simple convergence criterion here. We could, of course use other detection criteria, such as setting a probability threshold or looking at the differences between possible evolution of trajectories in the presence or absence of target objects. 

The left panel of Fig. \ref{fig:trajectories} shows a set of trajectories for the posterior probability of an object being present as a function of the number of shots, when a target object is actually present so the detection statistics at the monitoring detector are determined by the presence of this target. The three trajectories shown are the coherent state (pink, lowest trace), the heralded TMSV (light blue, middle trace) and click-matched heralded TMSV (orange, top trace). Each is produced from the same set of random numbers for the monitoring detector and the two heralded traces have the same set of random numbers for the idler detector. The traces show a significant amount of noise but it seems clear that the heralded traces provide a much more stable, quicker detection and that it is better to click match. The smoother curves in orange, navy and pink are the averages of 3000 trajectories. As an example of the utility of heralded TMSV states we use these to examine when the probability of object presence passes 0.8. The TMSV does this in less than half the number of shots of the coherent state and the click-matched TMSV in about a fifth of the number of shots.

The right panel in Fig. \ref{fig:trajectories} shows the object present probability traces, this time when a target is not present to determine the monitoring detector statistics. The same click count distribution produces all traces, but the fact that which signal state was sent is known in each case allows a different updating of the probability. Similar advantages to the left panel are shown in excluding the presence of the object for the heralded TMSV and the click-matched heralded TMSV. 

\section{Conclusions}

In this paper we have described a theory of quantum illumination for target detection in a noisy background. Our theory is written in terms of the the formalism of Gaussian quantum optics and can be wholly characterized using thermal Gaussian states, even in cases where the heralded state has a negative Wigner function and cannot be written as a Gaussian state. The theory is frequency-independent, so applies equally-well to lidar in the optical frequency range and to radar at microwave or radio wave frequencies, although the technical challenges of detectors sensitive to single photons are an issue here, as is the background rate at which they would fire. Research on such detectors is ongoing \cite{Inomata2016,Romero2009,Cridland2016}.

We have used our theory to show that the use of quantum illumination to provide a click-heralded state of TMSV of low average signal mean photon number provides a clear advantage, compared to using coherent states of the same mean photon number, for object detection under lossy, high background noise conditions. The return signal, under certain scenarios, is shown to be significantly more distinguishable from background noise. It provides enhanced click probabilities and, in turn, enhanced posterior probabilities useful for hypothesis testing in the multi-shot scenario. The detection of objects based on click-counts is also more stable definitive.

The quantum illumination advantage is a direct consequence of the photon number correlations between two spatially separate modes of the TMSV, even though each individual beam has mean photon number $\nbar$. If the heralding detector has high quantum efficiency and a low mean dark count probability, characterized by $\nbar_d$, the average increase in the mean photon number of the signal beam $\nbar_s$ is 1. This is a much more prominent effect for $\hat{\rho}_{\textsc{vst}}$ produced from a low average photon number TMSV. When the idler clicks, the photon number distribution in the signal arm shifts away from the vacuum leading to higher click probability. This contrasts with the coherent state which becomes more vacuum-like at low mean photon number. 

At higher mean photon number the heralding effect is much smaller and the coherent state provides better discrimination than the heralded TMSV. This is because the heralding has less effect on the TMSV and the coherent state probability distribution is more sharply-peaked around its mean value. We can, however, recover and increase the advantage by using a heralded TMSV with a higher unheralded mean photon number. This may appear to be cheating, but it can be accomplished by matching the detector click probabilities of the TMSV and coherent signal states, rendering them effectively indistinguishable to a Geiger-mode detector. This extra advantage might more correctly be termed a thermal state advantage, as it would also exist for illumination with classical single mode thermal states, but the combination of this and the heralding means that the quantum illumination can always outperform classical illumination.
 
Immediate advantages of quantum illumination using click-detectors are the readily available cost-effective equipment applicable to lidar systems, as well as he possibility of covert enhancement of photon number. Yet, the main drawback for performing quantum illumination experimentally would be $\hat{\rho}_\textsc{vst}$ production at the low $\nbar$ regime, as heralding probability for the TMSV dwindles quickly to zero via $ \frac{\nbar}{1+\nbar}$. Ideally we would require a reliable entangled source, so that we could run the heralding process at high frequency to provide a sufficient rate of heralded quantum state production. The output of a laser is much easier to use for illumination. There are, however, other advantages to using quantum illumination: we have more control over our state. The optical field has more degrees of freedom to be exploited: spatial, timing and polarisation to name but three. We shall explore measurement conditioned state-engineering with multiple click-detectors in future works, as this would provide an interesting crossover between basic click detection and photon number discrimination schemes with quantum illumination \cite{Ortolano2020, Sperling2014a}.

\section*{Funding}
The authors would like to thank the following funders: the UK Engineering and Physical Sciences Research Council for funding via the UK National Quantum Technology Programme and the QuantIC Imaging Hub (Grant Number EP/T00097X/1), QinetiQ and the University of Strathclyde.

\section*{Disclosures}
The authors declare no conflicts of interest.

\section*{Data availability}
Numerical data produced for figures are openly available from
the University of Strathclyde KnowledgeBase  \hyperlink{https://doi.org/10.15129/dbfd57cf-b189-41dc-9e99-ee25d172c3c7 (2021)}{here}.

\bibliography{library.bib}

\end{document}